\documentclass[12pt]{elsarticle}
\usepackage{amssymb}
\usepackage{amsmath}
\usepackage{fullpage}
\newcommand{\scA}{n{\cdot}(\partial{\wedge} A)}
\newcommand{\scB}{n{\cdot}(\partial{\wedge} B)}
\newcommand{\bea}{\begin{eqnarray}}
\newcommand{\eea}{\end{eqnarray}}
\newcommand{\nn}{\nonumber}

\begin{document}

\begin{frontmatter}
\title{SL(2,\textbf{R}) Duality-symmetric Action for Electromagnetic Theory with Electric and Magnetic Sources}
\author[rvt]{Choonkyu Lee} \ead{cklee@phya.snu.ac.kr}
\address[rvt]{Department of Physics and Astronomy and Center for Theoretical Physics\\ Seoul National University, Seoul 151-747, Korea}
\author[rvt1]{Hyunsoo Min} \ead{hsmin@dirac.uos.ac.kr}
\address[rvt1]{Department of Physics, University of Seoul, Seoul 130-743, Korea}

\begin{abstract}
For the SL(2,\textbf{R}) duality-invariant generalization of Maxwell electrodynamics in the presence of both electric and magnetic sources, we formulate a local, manifestly duality-symmetric, Zwanziger-type action by introducing a pair of four-potentials $A^\mu$ and $B^\mu$ in a judicious way. On the two potentials $A^\mu$ and $B^\mu$ the SL(2,\textbf{R}) duality transformation acts in a simple linear manner. In quantum theory including charged source fields, this action can be recast as a
 SL(2,\textbf{Z})-invariant action.
Also given is a Zwanziger-type action for SL(2,\textbf{R}) duality-invariant Born-Infeld electrodynamics which can be important for D-brane dynamics in string theory.
\end{abstract}
\end{frontmatter}

\newpage
\section{Introduction}
In recent days, duality has proved to be one of the most fruitful ideas in theoretical physics. Electromagnetic duality is the oldest and most studied example. Using covariant notations, Maxwell's equations with electric and magnetic sources\footnote{We assume the metric with signature $(-+++)$, and 
$\ ^{*\!}F^{\mu\nu}\equiv\frac{1}{2}\epsilon^{\mu\nu\lambda\delta}F_{\lambda\delta}$ (with $\epsilon^{0123}=1$) denotes the Hodge dual of an antisymmetric tensor $F^{\mu\nu}$. Note that $\ ^{*\!}(^{*\!}F^{\mu\nu})=-F^{\mu\nu}$.}
\begin{subequations}
\bea
\partial_\nu F^{\mu\nu}=J^\mu_{(e)}, \label{eq1a} \\
\partial_\nu \, ^{*\!}F^{\mu\nu}=J^\mu_{(m)} \label{eq1b}
\eea
\end{subequations}
and the related Lorentz 4-force
\bea
{\cal F}^\mu=\int d^3x \left[ F^{\mu\nu}(\vec{x},t)J_{(e)\nu}(\vec{x},t)+
\, ^{*\!}F^{\mu\nu}(\vec{x},t)J_{(m)\nu}(\vec{x},t)\right]  \label{eq2}
\eea
are invariant under the SO(2) duality rotation
\begin{subequations}
\bea
F^{\mu\nu}&\rightarrow& F^{\prime \mu\nu}=\cos\alpha\, F^{\mu\nu}
+\sin\alpha \, \,^{*\!}F^{\mu\nu}, \label{eq3a}\\
\left(\begin{array}{l}J^\mu_{(e)} \\ J^\mu_{(m)}\end{array}\right)&\rightarrow&
\left(\begin{array}{l}J^{\prime\mu}_{(e)} \\ J^{\prime\mu}_{(m)}\end{array}\right)=\left(\begin{array}{cc}\cos\alpha & \sin\alpha \\ -\sin\alpha &\cos\alpha \end{array}\right)\left(\begin{array}{l}J^\mu_{(e)} \\ J^\mu_{(m)}\end{array}\right). \label{eq3b}
\eea
\end{subequations}
Here, duality is not a symmetry in the usual sense but refers to the ambiguity in the theoretical description: what one calls electric versus magnetic is a matter of choice. With quantum effects taken into consideration, two additional developments have entered: (i) electric and magnetic charges must satisfy the Dirac quantization condition \cite{cit1} or, more generally, the Schwinger-Zwanziger condition \cite{cit2}, and (ii) the Maxwell Lagrangian can be modified by including a term proportional to 
$\theta \ ^{*\!}F^{\mu\nu}F_{\mu\nu}$,  by which a magnetic monopole acquires  a $\theta$-dependent electric charge \cite{cit3}. Because of these, the electromagnetic duality `symmetry' takes the form of the SL(2,\textbf{Z}) duality which can provide a strong constraint on quantum physics \cite{cit4}.

 If the above $\theta$-parameter is allowed to be spacetime-dependent or elevated to a dynamical field, we obtain the so-called axion electrodynamics \cite{cit5}; this theory has been studied not only from the perspective of particle physics \cite{cit6} but also from that of condensed matter physics \cite{cit7}. 
 Also  linear or nonlinear electromagnetic theories with maximal duality symmetry, the SL(2,\textbf{R}) invariance, have been addressed \cite{cit8,cit9}. To have such extended duality symmetry, additional scalar fields---dilation ($\phi$) and axion ($a$)---need to be introduced. For a linear system with SL(2,\textbf{R}) invariance, the Maxwell equations are generalized to
 \begin{subequations}
 \bea
 \partial_\nu\, ^{*\!}G^{\mu\nu}&=&J^\mu_{(m)},  \label{eq4a}\\
\partial_\nu H^{\mu\nu}&=&J^\mu_{(e)}, \qquad (H^{\mu\nu}(x)\equiv e^{-\phi(x)}G^{\mu\nu}(x)-a(x) ^{*\!}G^{\mu\nu}(x)) \label{eq4b}
\eea\end{subequations}
with the Lorentz force law now given by
\bea
{\cal F}^\mu=\int d^3x \left[ G^{\mu\nu}(\vec{x},t)J_{(e)\nu}(\vec{x},t)+
\,^{*\!}H^{\mu\nu}(\vec{x},t)J_{(m)\nu}(\vec{x},t)\right]. \label{eq5}
\eea
This theory is invariant under the duality transformation of the form
\begin{subequations}
\bea
\left(\begin{array}{l}G^{\mu\nu}\\ \, ^{*\!}H^{\mu\nu}\end{array}\right)\rightarrow \left(\begin{array}{l}G^{\prime\mu\nu}\\ ^{*\!}H^{\prime\mu\nu}\end{array}\right) =\left(\begin{array}{cc}s &\  r \\ q &\ p \end{array}\right)\left(\begin{array}{l}G^{\mu\nu}\\ ^{*\!}H^{\mu\nu}\end{array}\right)  \label{eq6a}
\eea
\bea
\tau\rightarrow \tau^\prime =\frac{p\tau+q}{r\tau +s},   \label{eq6b}
\eea
\bea
\left(\begin{array}{l}J^\mu_{(e)} \\ J^\mu_{(m)}\end{array}\right)\rightarrow
\left(\begin{array}{l}J^{\prime\mu}_{(e)} \\ J^{\prime\mu}_{(m)}\end{array}\right)=\left(\begin{array}{cc}p &-q \\-r & s\end{array}\right)\left(\begin{array}{l}J^\mu_{(e)} \\ J^\mu_{(m)}\end{array}\right),\label{eq6c}
\eea
\end{subequations}
 where we introduced a complex scalar
 \bea
 \tau(x)\equiv a(x)+ i\, e^{-\phi(x)},  \label{eq7}
\eea
and $p$, $q$, $r$, $s$ are real numbers satisfying the condition $ps-qr=1$, i.e., {\scriptsize$ \left({\begin{array}{cc}p &q \\r & s\end{array}}\right)$}$\in$ SL(2,\textbf{R}). With $\phi(x)$ set to zero, we obtain the equations of motion for axion electrodynamics from the system (\ref{eq4a})-(\ref{eq4b}). In fact, if the Maxwell theory is generalized to have spacetime-dependent $\theta$ and also spacetime-dependent electromagnetic coupling $e$ as in the Janus-type theory \cite{cit10}, the corresponding equations of motion (for rescaled fields) may also be put into the form (\ref{eq4a})-(\ref{eq4b}) with $\tau(x)=\frac{\theta(x)}{2\pi}+i\frac{4\pi}{e^2(x)}$ . In addition, we note that, for nonlinear Born-Infeld electrodynamic theory \cite{cit11} which finds an application as the world volume theory of a D-brane in string theory \cite{cit12}, its 
SL(2,\textbf{R})-invariant generalization has been obtained in Ref. \cite{cit8}. It is expected that, in the future theoretical development, this extended duality has a further important role.

But there is a curious point---while the duality symmetry is easily stated
with the equations of motion, it is not trivial to have it manifest at the action level.
This is because formulating the explicitly duality-symmetric action 
(with or without the sources present)
 is not quite trivial, especially when one looks for a local action based on a small number of independent potential variables. As for the Maxwellian system defined by (\ref{eq1a})-(\ref{eq1b}), Zwanziger  \cite{cit13} was able to formulate a relatively simple, SO(2) duality-invariant local QED action based on two four-potentials. (Another formalism was given by Schwarz and Sen \cite{cit14}, on which we make a brief comment in Sec.IV).
   The Zwanziger action has then become a basis for various quantum field-theoretic studies involving dyons \cite{cit15,cit16}. But, for the SL(2,\textbf{R})-invariant system specified by (\ref{eq4a})-(\ref{eq4b}), no such action has been put forward.\footnote{In the special dilaton-free case with $a(x)=\mathrm{const.}$, the authors of Ref. \cite{cit17} discussed the action form equivalent to ours after an appropriate SL(2,\textbf{R}) transformation. On this, see the remark after (\ref{eq46}).} 
  In this paper, we shall fill in this gap by providing a local, manifestly SL(2,\textbf{R})-invariant, Zwanziger-type action with both electric and magnetic sources. The method employed by us to derive this action proves to be applicable to a nonlinear electrodynamic theory also, as is ascertained from our construction of a local, SL(2,\textbf{R})-invariant, action based on Zwanziger-type four-potentials for the Born-Infeld theory by the same strategy. Existence of the local action formalism with maximal duality symmetry should be of some value in studying for instance quantum aspects of the theory.

 The rest of this paper is organized as follows. In Sec.II, a systematic formulation of the local Zwanziger-type action for the SL(2,\textbf{R})-invariant Maxwell-like system of (\ref{eq4a})-(\ref{eq4b}) is given. We here explain how Zwanziger-type four-potentials should be introduced in this case, and then go on to construct the desired local action based on those potentials variables by rewriting the related first-order nonlocal action for the system. The action obtained in this manner possesses manifest SL(2,\textbf{R}) invariance. In the corresponding quantum 
 theory including charged source fields, incorporating the charge quantization condition reduces the duality symmetry group to SL(2,\textbf{Z}) and our action can then be recast as a manifestly
 SL(2,\textbf{Z})-invariant action.
In Sec III, similar procedures are used to obtain the invariant local action for the SL(2,\textbf{R})-invariant Born-Infeld system of Ref.\cite{cit8}. (Only the classical action will be
considered here). In Sec IV, we summarize our work and offer some discussions related to our findings.

\section{SL(2,\textbf{R})-Invariant Action For The Maxwellian System}

 For the system with the equations of motion (\ref{eq1a})-(\ref{eq1b}), Zwanziger \cite{cit13} constructed a duality-invariant local action
 \bea
 S=S_Z(A,B)+\int d^4x[J^\mu_{(e)}A_\mu+ J^\mu_{(m)}B_\mu], \label{eq8}
 \eea
where  $A^\mu$, $B^\mu$ are two (independent) four-potentials in terms of which the Maxwell field strengths $F^{\mu\nu}\equiv (F)^{\mu\nu}$ can be expressed as\footnote{We
use Zwanziger's shorthand notations of writing $a^\mu b_\mu\equiv a{\cdot} b$,
$a^\mu a_\mu\equiv a^2$, $a^\mu b^\nu-a^\nu b^\mu\equiv(a{\wedge} b)^{\mu\nu}$,  $\partial^\mu a^\nu-\partial^\nu a^\mu\equiv(\partial{\wedge} a)^{\mu\nu}$, and $a_\mu F^{\mu\nu}\equiv(a{\cdot} F)^\nu$ for four vectors $a$ and $b$ and antisymmetric tensors $F^{\mu\nu}$. Note that $^{*\!}[a{\wedge} b]^{\mu\nu}\equiv \frac{1}{2} \epsilon^{\mu\nu}_{\quad \lambda\delta}(a{\wedge} b)^{\lambda\delta}=\epsilon^{\mu\nu}_{\quad\lambda\delta}a^\lambda b^\delta$.  }
\bea
F=n{\wedge}[n{\cdot}(\partial{\wedge} A)]- \, ^{*\!}\{n{\wedge}[n{\cdot}(\partial{\wedge} B)]\}  \label{eq9}
\eea
($n^\mu$ is a fixed, spacelike, four-vector, which is assumed to be normalized, i.e., $n^\mu n_\mu=1$). Under the duality transformation (\ref{eq3a})-(\ref{eq3b}) the potentials make an SO(2) doublet, that is, 
\bea
\left(\begin{array}{l}A^\mu \\ B^\mu\end{array}\right)\rightarrow
\left(\begin{array}{l}A^{'\mu}\\ B^{'\mu}\end{array}\right)=\left(\begin{array}{cc}\cos\alpha & \sin\alpha \\ -\sin\alpha &\cos\alpha \end{array}\right)\left(\begin{array}{l}A^\mu \\ B^\mu\end{array}\right).  \label{eq10}
\eea
$S_Z(A,B$) in (\ref{eq8}) has the SO(2) duality-invariant form
\bea
S_Z(A,B)&=&-\frac{1}{2}\int d^4 x\Big\{[n{\cdot}(\partial{\wedge} A)]{\cdot}[n{\cdot}\, ^{*\!}(\partial{\wedge} B)]\qquad  \nn \\
&&- [n{\cdot}(\partial{\wedge} B)]{\cdot}[n{\cdot}\, ^{*\!}(\partial{\wedge} A)]+[n{\cdot}(\partial{\wedge} A)]^2
+[n{\cdot}(\partial{\wedge} B)]^2  \Big\}. \label{eq11}
\eea
For the source currents $J^\mu_{(e)}$ and $J^\mu_{(m)}$, one may assume the form
\bea
J^\mu_{(e)}=\sum_j q_j \int ds_j \frac{d z^\mu_j(s_j)}{d s_j}\delta^4(x-z_j(s_j)), \nn \\
J^\mu_{(m)}=\sum_j g_j \int ds_j \frac{d z^\mu_j(s_j)}{d s_j}\delta^4(x-z_j(s_j)) \label{eq12}
\eea
considering point particles with electric and magnetic charges ($q_j, g_j$) ($z_j(s_j)$ denote their trajectories), or 
\bea
J^\mu_{(e)}=\sum_j q_j\overline{\psi}_j(x)\gamma^\mu\psi_j(x),\quad 
J^\mu_{(m)}=\sum_j g_j\overline{\psi}_j(x)\gamma^\mu\psi_j(x), \label{eq13}
\eea
if one has in mind the minimal coupling to a set of Dirac fields with appropriate electric and magnetic charges ($q_j, g_j$).
(Note that the SL(2,\textbf{R}) transformation of the charges $(q_j,g_j)$ is dictated by the
source transformation law (\ref{eq6c})).
 But, for classical discussion concentrating on Maxwell dynamics, there is no need to be specific as regards the nature of our sources. We also remark that, even if there enter two potentials $A^\mu$ and $B^\mu$ in the action (with U(1) gauge symmetry for each), we have only one photon in this theory \cite{cit13,cit15}.
 
 We want to find SL(2,\textbf{R})-invariant action for the system described by the equations of motion (\ref{eq4a})-(\ref{eq4b}), as an extension of the above approach. For this, two things will have to be answered: with the new equations of motion, 
\begin{enumerate}
\item[1.] What would be the correct way of introducing the two four-potentials $A^\mu$ and $B^\mu$ to represent the antisymmetric field strengths $G^{\mu\nu}$? (That is, what is the formula replacing (\ref{eq9})?), and
\item[2.] then, based on the two four-potentials, what would be the appropriate action form (replacing (\ref{eq11}) and exhibiting the manifest SL(2,\textbf{R}) symmetry)?
\end{enumerate}
Note that our action will have the same source-dependent terms as in (\ref{eq8}); then, from the SL(2,\textbf{R}) invariance and the source transformation law (6c), our potentials $A^\mu$ and $B^\mu$ must obey the SL(2,\textbf{R}) transformation law
\bea
\left(\begin{array}{l}A^\mu \\ B^\mu\end{array}\right)\rightarrow
\left(\begin{array}{l}A^{'\mu}\\ B^{'\mu}\end{array}\right)=\left(\begin{array}{cc}s&\ r\\q&\ p\end{array}\right)\left(\begin{array}{l}A^\mu \\ B^\mu\end{array}\right). \label{eq14}
\eea
(Note that {\scriptsize$\left(\begin{array}{cc}p&-q\\-r&s\end{array}\right)^{\!\mathrm{T}}\left(\begin{array}{cc}s&\ r\\q&\ p\end{array}\right)=\left(\begin{array}{cc}1&\ 0\\0&\ 1\end{array}\right)$}).
Keeping this in mind, we shall below provide the answers to the above questions. In our consideration that follows, the two scalars  $\phi(x)$ and $a(x)$  assume only passive roles and so will be treated as if they correspond to externally given backgrounds. If one wishes, the SL(2,\textbf{R})-invariant `kinetic' action for the scalars (given in Ref. \cite{cit8}) as well as the appropriate source kinetic action may of course be included in our final action.

 The right formula expressing $G^{\mu\nu}$ in terms of the potentials $A^\mu$ and $B^\mu$ can be found as follows. Note that, if (\ref{eq4a}) holds (and hence 
 $\partial{\cdot} J_{(m)}=0$), it is always possible to introduce the four-potential $A^\mu$ such that
 \bea
 G^{\mu\nu}(x)=(\partial{\wedge} A)^{\mu\nu}(x)+\int d^4x' \,^{*\!}[f(x-x'){\wedge}
 J_{(m)}(x')]^{\mu\nu}  \label{eq15}
 \eea
where $f^\mu(x-x')$ is any vector function satisfying
\bea
\partial_\mu f^\mu(x-x')=\delta^4(x-x').   \label{eq16}
\eea
Following Zwanziger \cite{cit13}, we will choose $f^\mu(x-x')=n^\mu(n{\cdot} \partial)^{-1}(x,x')$
with $(n{\cdot} \partial)^{-1}(x,x')$ having the explicit representation
\footnote{We consider only the theory which is duality invariant. For this reason we have
chosen our kernel $(n{\cdot}\partial)^{-1}$ to contain strings in a symmetrical way  \cite{cit13,cit15}.}
\bea
(n{\cdot} \partial)^{-1}(x,x')&=&\frac{1}{2}\{\theta[n{\cdot}(x-x')]-
\theta[-n{\cdot}(x-x')]\}\delta_n^3(x-x')  \label{eq17}\\ 
(&=&-(n{\cdot} \partial)^{-1}(x',x)), \nn
\eea
so that the spacelike unit vector $n^\mu$ may correspond to the direction of the Dirac string. ($\delta_n^3(x-x')$ in (\ref{eq17}) denotes the 3-dimensional $\delta$-function defined on the hypersurface orthogonal to $n^\mu$). We may thus write
\bea
G^{\mu\nu}=(\partial{\wedge} A)^{\mu\nu}+(n{\cdot}\partial)^{-1}\, ^{*\!}(n{\wedge} J_{(m)})^{\mu\nu}. \label{eq18}
\eea
Similarly, from (4b), we can consider another four-potential $B^\mu$ such that 
\bea
H^{\mu\nu}\equiv e^{-\phi}G^{\mu\nu}- a\, ^{*\!}G^{\mu\nu}=
-^{*\!}(\partial{\wedge} B)^{\mu\nu}-(n{\cdot}\partial)^{-1}(n{\wedge} J_{(e)})^{\mu\nu}. \label{eq19}
\eea
From these two relations follow
\begin{subequations}
\bea
n{\cdot} G&=& n{\cdot} (\partial{\wedge} A),   \label{eq20a}\\
n{\cdot}\, ^{*\!}G&=&e^\phi [n{\cdot}(\partial{\wedge} B)-a \, n{\cdot}(\partial{\wedge} A)]. \label{eq20b}
\eea
\end{subequations}
If we use these results with the Zwanziger identity (valid for any antisymmetric tensor $Q_{\mu\nu}$)
\bea
Q=n{\wedge}(n{\cdot} Q)-\, ^{*\!}[n{\wedge}(n{\cdot}\,^{*\!}Q)], \label{eq21}
\eea
we will be led to the expression of $G^{\mu\nu}$ in terms of the potentials we introduced:
\bea
 G= n{\wedge}[n{\cdot} (\partial{\wedge} A)]
-e^\phi\  ^{*\!}\{n{\wedge} [n{\cdot}(\partial{\wedge} B)-a\, n{\cdot}(\partial{\wedge} A)]\}. \label{eq22}
\eea
It turns out that this is the formula we sought. Then, for $^{*\!}G^{\mu\nu}$, we have
\bea
^{*\!}G=
e^\phi n{\wedge} [n{\cdot}(\partial{\wedge} B)-a\, n{\cdot}(\partial{\wedge} A)]+
\, ^{*\!}\{n{\wedge}[n{\cdot} (\partial{\wedge} A)]\},  \label{eq23}
\eea
and accordingly, for the tensor $H^{\mu\nu}\equiv (H)^{\mu\nu}$ defined in (4b),
\bea
H=e^\phi (a^2+e^{-2\phi})n{\wedge}[n{\cdot}(\partial{\wedge} A)]-\, ^{*\!}\{n{\wedge}[n{\cdot}(\partial{\wedge} B)]\}
-a \, e^\phi n{\wedge}[n{\cdot}(\partial{\wedge} B)].  \label{eq24}
\eea

To justify the connection (\ref{eq22}), we are obliged to check the SL(2,\textbf{R}) transformation property for the potentials $A^\mu$ and $B^\mu$ defined through (\ref{eq22}) for given $G^{\mu\nu}$. Here let us define the complex field strengths ${\cal G}^{\mu\nu}\equiv
({\cal G})^{\mu\nu}$ by
\bea
{\cal G}= G+ i \, ^{*\!}G .  \label{eq25}
\eea
This allows us to describe the SL(2,\textbf{R}) transformation (\ref{eq6a}) by the formula
\bea
{\cal G}\rightarrow {\cal G}'=(r \bar{\tau} +s ) {\cal G} \label{eq26}
\eea
($\bar{\tau}$ is the complex conjugate of $\tau$), which may be used together with the 
M\"obius transformation (\ref{eq6b}) for $\tau$. On the other hand, if our form (\ref{eq22}) is used for 
$G$, the complex field strengths would acquire the form
\bea
{\cal G}= e^\phi\  \bigg(\!-i n{\wedge}[n{\cdot} (\tau \partial {\wedge} A-\partial {\wedge} B)] 
+\, ^{*\!}\Big\{n{\wedge}[n{\cdot}(\tau \partial {\wedge} A- \partial {\wedge} B)]\Big\}\bigg). \label{eq27}
\eea
Then, because of the transformation law (\ref{eq26}), the SL(2,\textbf{R}) transformation law for the potential $A^\mu$ and $B^\mu$ would have to be such that
\bea
n{\cdot} (\tau \partial {\wedge} A-\partial {\wedge} B)\rightarrow 
n{\cdot} (\tau' \partial {\wedge} A'-\partial {\wedge} B')=
\frac{1}{r \tau+s} n{\cdot} (\tau \partial {\wedge} A-\partial {\wedge} B), \label{eq28}
\eea
since, from the M\"obius transformation (\ref{eq6b}), 
\bea
e^{-\phi}\rightarrow e^{-\phi'}=\frac{e^{-\phi}}{(r \bar{\tau}+s)(r\tau+s)}.  \label{eq29}
\eea
It takes a  short calculation to verify that the law (\ref{eq28}) is realized if the potentials
 $A^\mu$ and $B^\mu$ follow the SL(2,\textbf{R}) transformation law as given by (\ref{eq14}). (Actually
 it suffices only if the curls $\partial{\wedge} A$ and $\partial{\wedge} B$ follow the 
 corresponding transformation law deduced from (\ref{eq14}), for (\ref{eq28}) to be true).
 
To find the appropriate  action having the potentials $A^\mu$ and $B^\mu$ as independent
variables, it is useful  to start from the nonlocal, first-order, action for our system
with $^{*\!}G^{\mu\nu}$($=-^{*\!}G^{\nu\mu} $) and $B^\mu$ taken as variables. As such action form, consider the expression
\bea
S&=&\int d^4x \left[\frac{1}{4} e^{-\phi}\, ^{*\!}G^{\mu\nu}\, ^{*\!}G_{\mu\nu}
-\frac{1}{2}\, ^{*\!}G^{\mu\nu}(\partial_\mu B_\nu-\partial_\nu B_\mu)+\frac{1}{4}
a  \, ^{*\!}G^{\mu\nu}\, G_{\mu\nu}\right] \nn \\
&&+\int d^4 x\left[ J^\mu_{(e)}A_\mu(G)+J^\mu_{(m)} B_\mu\right],  \label{eq30}
\eea
where $G^{\mu\nu}=-\, ^{*\!}(^{*\!}G)^{\mu\nu}$ and $A_\mu(G)$ denotes the nonlocal function of $G$ specified by
\bea
A_\mu=(n{\cdot}\partial)^{-1}(n{\cdot} G)_\mu.  \label{eq31}
\eea
(This nonlocal action can be looked upon as the generalization of the Schwinger-Yan-type 
action [18] for the system (\ref{eq1a})-(\ref{eq1b}) to our case). With the action (\ref{eq30}), variation of the
independent variables $B_\mu$ and $^{*\!}G_{\mu\nu}$ yields the equations of motion
\begin{subequations}
\bea
\partial_\nu \,^{*\!}G^{\mu\nu}&=&J^\mu_{(m)},  \label{eq32a}\\
e^{-\phi}\, ^{*\!}G^{\mu\nu}+a G^{\mu\nu}&=&\partial^\mu B^\nu -\partial^\nu B^\mu
-(n{\cdot}\partial)^{-1} \, ^{*\!}(n{\wedge} J_{(e)})^{\mu\nu},  \label{eq32b}
\eea
\end{subequations}
where, for the  second equation, the property $(n{\cdot}\partial)^{-1}(x,x')=
-(n{\cdot}\partial)^{-1}(x',x)$ (see (\ref{eq17})) has been used. (\ref{eq32b}) is nothing but the dual of (\ref{eq19}). Hence the equations of motion comprise of (\ref{eq4a}) and (\ref{eq19}), but (4b) is satisfied if the relation (\ref{eq19}) is given. Further, from (\ref{eq31}) and (\ref{eq32a}), one can readily show that
our relation (\ref{eq18}) is also valid. Therefore, we may consistently use the representation (\ref{eq23})
for $^{*\!}G^{\mu\nu}$ inside the above action and regard the potential $B^{\mu}$ and $A^\mu$ as independent variables.

The integrand  of the first part in the right hand side of (\ref{eq30}),
\bea
&&\frac{1}{4} e^{-\phi} \, ^{*\!}G^{\mu\nu}\, ^{*\!}G_{\mu\nu}
-\frac{1}{2}\, ^{*\!}G^{\mu\nu}(\partial_\mu B_\nu-\partial_\nu B_\mu)+\frac{1}{4}
a  \, ^{*\!}G^{\mu\nu}\, G_{\mu\nu} \qquad \qquad \nn \\
&&\quad=\frac{1}{4}\,  ^{*\!}G^{\mu\nu}\, ^{*\!}H_{\mu\nu}-\frac{1}{2}\,  ^{*\!}G^{\mu\nu}(\partial^\mu B^\nu -\partial^\nu B^\mu),  \label{eq33}
\eea
can readily be changed into an expression involving the potentials by making use of our
representations (\ref{eq23}) and (\ref{eq24}).  This way, we obtain the action that has the potentials 
 $A^\mu$ and $B^\mu$ as variables:
 \bea
 S&=&-\frac{1}{2}\int d^4 x \Big\{ [n{\cdot}(\partial{\wedge} A)]{\cdot}[n{\cdot}\, ^{*\!}(\partial{\wedge} B)]- [n{\cdot}(\partial{\wedge} B)]{\cdot}[n{\cdot}\, ^{*\!}(\partial{\wedge} A)] \nn \\
 &&\qquad \qquad\quad+ e^\phi(a^2+e^{-2\phi})[n{\cdot}(\partial{\wedge} A)]^2+e^\phi[n{\cdot}(\partial{\wedge} B)]^2 \nn \\
&&\qquad -2a e^\phi [n{\cdot}(\partial{\wedge} A)]{\cdot} [n{\cdot}(\partial{\wedge} B)]\Big\} 
+\int d^4 x\left[ J^\mu_{(e)}A_\mu+J^\mu_{(m)} B_\mu\right]. \label{eq34}
\eea
Notice that the old Zwanziger action, given by (\ref{eq8}) and (\ref{eq11}), is  recovered from this expression upon setting $\phi=a=0$. Using the complex scalar $\tau$, it is possible
to present our action (\ref{eq34}) by a simpler form of
\bea
S&=&\frac{1}{2}\int d^4x \Big\{ {\rm Im}
\left[e^\phi (n{\cdot} [\tau\partial{\wedge} A-\partial{\wedge} B ] ){\cdot}
(n{\cdot}\, ^{*\!}[\bar{\tau}\partial{\wedge} A-\partial{\wedge} B]) \right]  \nn \\
&&\qquad \qquad - {\rm Re}\left[ e^\phi 
(n{\cdot}[\tau\partial{\wedge} A-\partial{\wedge} B]){\cdot}
(n{\cdot}[\bar{\tau}\partial{\wedge} A-\partial{\wedge} B] ) \right] \Big\} \nn \\
&& +\int d^4 x\left[ J^\mu_{(e)}A_\mu+J^\mu_{(m)} B_\mu\right]. \label{eq35}
\eea
The SL(2,\textbf{R}) duality invariance is particularly evident in this latter form in view
of the transformations (\ref{eq28}) and (\ref{eq29}). Also, as one can verify explicitly, this action
leads to the correct form of equations of motion---viz., (\ref{eq4a}) with
$^{*\!}G^{\mu\nu}=\, ^{*\!}G^{\mu\nu}(A,B)$ as given by (\ref{eq23}) follows from the variation of $B_\mu$,
and (4b) with $H^{\mu\nu}=H^{\mu\nu}(A,B)$ as given by (\ref{eq24}) from the variation of $A_\mu$.

From that we have (\ref{eq22}) and (\ref{eq4a})-(\ref{eq4b}), can we assert the validity of the relations (\ref{eq18}) and (\ref{eq19}) also? This will be an important question for source dynamics (see below), and the answer is in the affirmative. For the validity of (\ref{eq18}), we may here demonstrate the relation
\bea
 ^{*\!}G-(n{\cdot}\partial)^{-1} n{\wedge}(\partial{\cdot}\, ^{*\!}G)=\,  ^{*\!}(\partial{\wedge} A), \label{eq36}
\eea
which is directly related to (\ref{eq18}) upon using (\ref{eq4a}). If the representation (\ref{eq23}) is used for $^{*\!}G$, we have in fact
\bea
(n{\cdot} \partial)\, ^{*\!}G^{\mu\nu}-[n{\wedge}(\partial{\cdot}^{*\!}G)]^{\mu\nu}=(n{\cdot} \partial)\, ^{*\!}(\partial{\wedge} A)^{\mu\nu}, \label{eq37}
\eea
thanks to the identity $\epsilon^{\lambda\nu\kappa\delta}n^\mu+\epsilon^{\nu\kappa\delta\mu}n^\lambda
+\epsilon^{\kappa\delta\mu\lambda}n^\nu+\epsilon^{\mu\lambda\nu\kappa}n^\delta+\epsilon^{\delta\mu\lambda\nu}n^\kappa=0$,
and (\ref{eq36}) follows from this relation as we assume the boundary condition
$^{*\!}G^{\mu\nu}\to 0$ for $n{\cdot} x \to \infty$. Similarly, for 
the validity of (\ref{eq19}), one may
utilize the following relation that holds for $H^{\mu\nu}$ given by (\ref{eq24}):
\bea
(n{\cdot}\partial )H^{\mu\nu}-[n{\wedge}(n{\cdot} H)]^{\mu\nu}=-(n{\cdot}\partial )
\, ^{*\!}(\partial{\wedge} B)^{\mu\nu}. \label{eq38}
\eea
Implications of (\ref{eq18})and (\ref{eq19}) are clear: outside the string locations, the curls
$(\partial{\wedge} A)^{\mu\nu}$ and $(\partial{\wedge} B)^{\mu\nu}$ are equal to the field
strengths $G^{\mu\nu}$ and $^{*\!}H^{\mu\nu}$, respectively. Based on this finding, we can 
now say something on source dynamics, especially on the Lorentz 
force structure. Because of the minimal coupling assumed, the Lorentz
4-force ${\cal F}^\mu\equiv \int d^3x \, \mathfrak{ f}^\mu (x)$ that one would find in association with the action (\ref{eq35}) should read
\bea
\mathfrak{ f}^\mu(x)=(\partial{\wedge} A)^{\mu\nu}(x) J_{(e)\nu}(x)+
(\partial{\wedge} B)^{\mu\nu}(x) J_{(m)\nu}(x). \label{eq39}
\eea
Then, using (\ref{eq18}) and (\ref{eq19}), this can be rewritten as
\bea
\mathfrak{ f}^\mu(x)&=&G^{\mu\nu}(x) J_{(e)\nu}(x)+
\, ^{*\!}H^{\mu\nu}(x) J_{(m)\nu}(x) \nn \\
&&+ \mbox{(terms vanishing outside the  string locations)}.  \label{eq40}
\eea  
We thus find (\ref{eq5}) from this result, if the term involving the (infinitesimally thin) string can be neglected.\footnote{Quantum mechanically, however, one finds physical predictions which are independent of the string  direction $n^\mu$ only when the appropriate quantization condition (see (\ref{eq48})) is fulfilled.}  Also evident from (\ref{eq18}) and (\ref{eq19}) is that,
in free space with $J_{(e)}=J_{(m)}=0$, our two potentials are related by
\bea
\partial{\wedge} B= e^{-\phi}\  ^{*\!}(\partial{\wedge} A) + a\ (\partial{\wedge} A), \label{eq41}
\eea
and therefore we have only one independent electromagnetic potential.

As for our action (\ref{eq34}) or (\ref{eq35}), there is another point to mention: when one specializes to the
case with {\em constant-valued scalars} $\phi$ and $a$, its SL(2,\textbf{R}) invariance may be utilized to obtain an
alternative, but physically equivalent, description of the given system. 
(The constant $a$ here may be related to the $\theta$ parameter, i.e., $a=\frac{e^2 \theta}{8\pi^2}$ where
$e$ denotes the electromagnetic coupling).
For any given constant complex
scalar $\tau=a+i e^{-\phi}$, the result of performing the transformation (\ref{eq6b}) with $p=e^{\phi/2}$, $q=-ae^{\phi/2}$, $r=0$ and $s=e^{-\phi/2}$ is $\tau'=i$, {i.e.}, leads to the values $a'=0$ and $\phi'=0$. Therefore, we may describe this system, using the SL(2,\textbf{R}) transformed potentials (see (\ref{eq14}))
\bea
A'^\mu=e^{-\phi/2} A^\mu, \quad B'^\mu=e^{\phi/2} (-a A^\mu + B^\mu)\label{eq42}
\eea
by the old Zwanziger action (see (\ref{eq11}))
\bea
S&=&-\frac{1}{2}\int d^4 x\Big\{[n{\cdot}(\partial{\wedge} A')]{\cdot}[n{\cdot}\, ^{*\!}(\partial{\wedge} B')]
- [n{\cdot}(\partial{\wedge} B')]{\cdot}[n{\cdot}\, ^{*\!}(\partial{\wedge} A')] \nn \\
&&\qquad \qquad\  +[n{\cdot}(\partial{\wedge} A')]^2
+[n{\cdot}(\partial{\wedge} B')]^2\Big\} \nn \\
&&+\int d^4x \left[ e^{\phi/2} (J^\mu_{(e)}+a J^\mu_{(m)})A'_\mu +e^{-\phi/2}J^\mu_{(m)} B'_\mu\right]. \label{eq43}
\eea
The source coupling structure is the only difference from Zwanziger's. For the source currents
$J^\mu_{(e)}$ and $J^\mu_{(m)}$ taken by the expression (\ref{eq12}) or (\ref{eq13}), this implies that the source
currents to which the new potentials $A'_\mu$ and  $B'_\mu$ couple come with effective electric and magnetic charges
\bea
e'=e^{\phi/2}(q+a g), \qquad g'=e^{-\phi/2} g \label{eq44}
\eea
for every particle/field possessing electric and magnetic charges $(q,g)$. (\ref{eq44}) corresponds to a (classical) generalization of the well-known Witten formula \cite{cit3} for the effective dyon electric charge.

There is a simple way to understand the above finding (with constant $\phi$ and $a$). In terms of the new
potentials in (\ref{eq42}), our field strengths $G^{\mu\nu}$ in (\ref{eq22}) assume the Zwanziger form (see (\ref{eq9})) up to a constant factor
\bea
G=e^{\phi/2}\Big( n{\wedge}[n{\cdot}(\partial{\wedge} A')]-\, ^{*\!}\{n{\wedge}[n{\cdot}(\partial{\wedge} B')]\}\Big),
\label{eq45}
\eea
while the equations of motion (\ref{eq4a})-(\ref{eq4b}) have the same physical content as
\begin{subequations}\label{eq46}
\bea
\partial_\nu \left( e^{-\phi/2}\, ^{*\!}G^{\mu\nu}\right)&=& e^{-\phi/2} J^\mu_{(m)} \qquad \qquad(\equiv  J'^\mu_{(m)}), \label{eq46a} \\
\partial_\nu \left( e^{-\phi/2}\, G^{\mu\nu}\right)&=& e^{\phi/2}( J^\mu_{(e)} + aJ^\mu_{(m)} ) \quad(\equiv  J'^\mu_{(e)}). \label{eq46b}
\eea\end{subequations}
Hence, if we define $F^{\mu\nu}=e^{-\phi/2}G^{\mu\nu}$, we find the SO(2) duality-invariant system of (\ref{eq1a})-(\ref{eq1b}), and so our action (\ref{eq43}). For the Maxwell dynamics in the presence of the $\theta$ term, 
the authors of Ref.[17] also proposed the action form comparable to our expression (\ref{eq43}).
[See eqs. (3.8) and (3.9) of Ref.[17]; actually, the action given  differs from our form (\ref{eq43}) by
a surface term, which had better be deleted according to our development]. 

We will now  discuss some aspects which are relevant for the development of quantized theory.
Our action (\ref{eq34}), with two 4-potential variables, contains more than the usual number of
redundant variables and so, to quantize this theory, one may have to resort to the quantization procedure appropriate to 
a system with constraints (as was  done by Balachandran {\it et al.}\cite{cit19} for the Zwanziger action (\ref{eq11})). Alternatively, following Zwanziger  \cite{cit13}, one may introduce a suitable gauge fixing term and then apply the canonical quantization scheme. In our case a natural SL(2,\textbf{R})-invariant gauge fixing term is provided by
\bea
S_{\mathrm{gf}}=\frac{1}{2} \int d^4 x\  \mathrm{Re}\Big[ e^\phi
\{\tau \partial^\mu(n{\cdot}A)-\partial^\mu(n{\cdot}B)\}
\{\bar{\tau}\partial_\mu(n{\cdot}A)-\partial_\mu(n{\cdot}B)\}\Big]. \label{eq47}
\eea
If this term is added to what we have in (\ref{eq34}), the resulting action (with the choice $n^\mu=(0, \hat{n})$) is
of first order in time derivative---it is a {\em phase space action} that may be quantized canonically.
This is the procedure taken in Ref.\cite{cit13}, and we expect no serious obstacle to arise in implementing this for
our case. From this canonical  analysis, one should also be able to see the reduction in relevant dynamical 
degrees of freedom, to those of a single photon.

Also, in connection with source field dynamics, one may repeat the analysis of Ref.\cite{cit13,cit15} to argue that electric and magnetic charges for the fields be such that the SL(2,\textbf{R})-invariant quantization condition \cite{cit2}
\bea
q_ig_j-q_jg_i= 4\pi n_{ij}, \qquad n_{ij}=0, \pm1, \pm2,\cdots \label{eq48}
\eea
may be fulfilled. This is a quantum consistency condition, needed as one demands that physical amplitudes should be independent
of the Dirac string direction (represented by $n^\mu$). Then, incorporating this condition (under the 
assumption that some fields are purely electrically charged), we may adjust our source currents in (\ref{eq13}) for instance to be given as
\bea
J^\mu_{(e)}(x)&=&e\sum_j(n_j+\frac{\theta_0}{2\pi} m_j) \bar{\psi}_j(x)\gamma^\mu\psi(x)\equiv
e[\tilde{J}^\mu_{(e)}(x)+\frac{\theta_0}{2\pi}\tilde{J}^\mu_{(m)}(x)],  \nn \\
J^\mu_{(m)}(x)&=&\frac{4\pi}{e}\sum_j m_j \bar{\psi}_j(x)\gamma^\mu\psi(x)
\equiv\frac{4\pi}{e}[\tilde{J}^\mu_{(m)}(x)],  \label{eq49}
\eea
($n_j$, $m_j$ are integers), with $e$ now interpreted as the coupling constant of the theory and $\theta_0$
a free, angle-like, parameter  \cite{cit20}. (In view of (\ref{eq44}), this free parameter $\theta_0$ is clearly interchangeable with the background axion field. See below.)
Once these current forms are assumed, the general SL(2,\textbf{R}) transformation in (\ref{eq6c}) is not
definable and hence the SL(2,\textbf{R})-invariant nature of our action (\ref{eq35}) would be of no particular significance. 
But, in quantum theory, we have SL(2,\textbf{Z}) duality \cite{cit4} instead, and by a simple reformulation based on our action (\ref{eq35}) it is possible to obtain a manifestly SL(2,\textbf{Z})-invariant action also. This is discussed below.

With the source currents given as in ({\ref{eq49}), it is convenient to introduce the new 4-potentials $\tilde{A}^\mu$
and $\tilde{B}^\mu$ by
\bea
\tilde{A}^\mu\equiv e A^\mu, \qquad \tilde{B}^\mu\equiv \frac{4\pi}{e}[ B^\mu+\frac{e^2 \theta_0}{8\pi^2} A^\mu].
\eea
Using these variables, the action (\ref{eq35}) with $a(x)\equiv \frac{e^2\theta(x)}{8\pi^2}$ can be
recast as
\bea
S&=&\frac{1}{2}\frac{1}{4\pi}\int d^4x \Big\{ {\rm Im}
\left[ e^{\tilde{\phi}} (n{\cdot} [\tilde{\tau}\partial{\wedge} \tilde{A}-\partial{\wedge} \tilde{B} ] ){\cdot}
(n{\cdot}\, ^{*\!}[\bar{\tilde{\tau}}\partial{\wedge} \tilde{A}-\partial{\wedge} \tilde{B}]) \right]  \nn \\
&&\qquad \qquad - {\rm Re}\left[e^{\tilde{\phi}} (n{\cdot} [\tilde{\tau}\partial{\wedge} \tilde{A}-\partial{\wedge} \tilde{B} ] ){\cdot}
(n{\cdot}\, [\bar{\tilde{\tau}}\partial{\wedge} \tilde{A}-\partial{\wedge} \tilde{B}]) \right]  \Big\} \nn \\
&& \qquad \qquad +\int d^4 x\left[ \tilde{J}^\mu_{(e)}\tilde{A}_\mu+\tilde{J}^\mu_{(m)} \tilde{B}_\mu\right]. \label{eq51} 
\eea
Here, $\tilde{\phi}(x)$ and $\tilde{\tau}(x)$ are defined by
\begin{subequations}
\bea
\tilde{\phi}(x)&=&\phi(x)+\phi_0, \quad (\mathrm{with} \quad e^{-\phi_0}\equiv\frac{4\pi}{e^2}),  \label{eq52a} \\
\tilde{\tau}(x)&=&e^{-\phi_0}\left\{\frac{e^2}{8\pi^2}[\theta(x)+\theta_0]+i\,e^{-\phi(x)} \right\} \nn \\
&=&\frac{\tilde{\theta}(x)}{2\pi}+ i\, e^{-\tilde{\phi}(x)}, \quad (\mathrm{with} \quad \tilde{\theta}(x)\equiv \theta(x)+\theta_0). \label{eq52b}
\eea
\end{subequations}
Using the variables with tilde, we thus find the action form having basically the same structure as the original form (\ref{eq35}).\footnote{Note that, in terms of the variables with tilde, the equations of motion (\ref{eq4a})-(\ref{eq4b})
have in fact the same content as
$\partial_\nu\, ^{*\!}\tilde{G}^{\mu\nu}=4\pi \tilde{J}^\mu_{(m)},  \ 
\partial_\nu[ e^{-\tilde{\phi}}\tilde{G}^{\mu\nu}(x)-\frac{\tilde{\theta}}{2\pi}\, ^{*\!}\tilde{G}^{\mu\nu}(x)]={4\pi}\tilde{J}^\mu_{(e)}$.}  
The field strengths $G^{\mu\nu}$ given by (\ref{eq22}) can also be expressed using the new variables in the structure-preserving manner:
\bea
G^{\mu\nu}=\frac{1}{e}\Big( n{\wedge}[n{\cdot}(\partial{\wedge}\tilde{A})]- e^{\tilde{\phi}}\, ^{*\!}\{n{\wedge}
[n{\cdot}(\partial{\wedge}\tilde{B})-\frac{\tilde{\theta}}{2\pi} n{\cdot}(\partial{\wedge}\tilde{A})]\}\Big)
\equiv \frac{1}{e}\tilde{G}_{\mu\nu}. \label{eq53}
\eea
Hence, in the quantum theory described using the variables with tilde, one might expect the same symmetry property 
to be present as in the classical case. But there is one condition--for 
$\tilde{J}^\mu_{(e)}(x)=\sum_j n_j \bar{\psi}_j(x) \gamma^\mu \psi_j(x)$ and 
$\tilde{J}^\mu_{(m)}(x)=\sum_j m_j \bar{\psi}_j(x) \gamma^\mu \psi_j(x)$ with integers $n_j$ and $m_j$, the constants
$p$, $q$, $r$, $s$ in the transformation laws (\ref{eq6a})-(\ref{eq6c}) and (\ref{eq14})
(now applied to the variables with tilde) will have to be restricted to integers satisfying $pq-rs=1$, {\it i.e.}
 {\scriptsize$ \left({\begin{array}{cc}p &q \\r & s\end{array}}\right)$}$\in$ SL(2,\textbf{Z}). 
This shows that, in quantum theory, we are left with the SL(2,\textbf{Z}) duality which is manifest in our action expression (\ref{eq51}).
Note that the SL(2,\textbf{Z}) transformation acts on the potentials $\tilde{A}^\mu$
and $\tilde{B}^\mu$ in a simple linear manner, { i.e.},
\bea  
\left(\begin{array}{c}\tilde{A}^\mu \\ \tilde{B}^\mu \end{array}\right)\rightarrow
\left(\begin{array}{c}\tilde{A'}^\mu \\ \tilde{B'}^\mu \end{array}\right)=\left({\begin{array}{cc}s &r \\q & p\end{array}}\right)\left(\begin{array}{c}\tilde{A}^\mu \\ \tilde{B}^\mu \end{array}\right). \label{eq54}
\eea

\section{SL(2,\textbf{R})-Invariant Born-Infeld Action}
Born-Infeld electrodynamics \cite{cit11} is a prototype nonlinear field theory exhibiting the SO(2) group of electric-magnetic duality symmetry \cite{cit21}. 
Gibbons and Rasheed \cite{cit8} extended this theory by including a coupling to axion ($a$) and dilaton ($\phi$) fields in such a way that the resulting theory may exhibit the SL(2,\textbf{R}) duality symmetry. 
 The equations of motion of the latter theory can still be written in the  form  (\ref{eq4a})-(\ref{eq4b}), i.e.,  as
 \begin{subequations}
 \bea
 \partial_\nu\, ^{*\!}G^{\mu\nu}&=&J^\mu_{(m)},  \label{eq4a2}\\
\partial_\nu H^{\mu\nu}&=&J^\mu_{(e)},  \label{eq4b2}
\eea\end{subequations}
 and  they share the same symmetry property
under the SL(2,\textbf{R}) duality transformation in (\ref{eq6a})-(\ref{eq6c}). 
However, the antisymmetric tensor $H^{\mu\nu}$ is now a highly nonlinear function of $G^{\mu\nu}$:
\bea
H^{\mu\nu}&=&\frac{ e^{-\phi}{G}^{\mu\nu}-\frac{1}{4} e^{-2\phi}(G{^{*\!}G})\,{^{*\!}G}^{\mu\nu}}
{\sqrt{1-\frac{1}{2}e^{-\phi}(^{*\!}G)^2-\frac{1}{16}e^{-2\phi}(G ^{*\!}G)^2}}- a\, ^{*\!}G^{\mu\nu} . \label{Hfield}
\eea
(We used the abbreviations, $(G\, ^{*\!}G)\equiv G^{\mu\nu}\, ^{*\!}G_{\mu\nu}$ and
$(\, ^{*\!}G)^2\equiv \, ^{*\!}G^{\mu\nu}\, ^{*\!}G_{\mu\nu}=-G^{\mu\nu}G_{\mu\nu}\equiv -(G)^2 $).
Note that this reduces to the expression in (\ref{eq4b}) if we expand (\ref{Hfield}) in a power series of $G^{\mu\nu}$ and keep only leading terms. As in the linear system discussed in the previous section, we want to find an expression for $G^{\mu\nu}$ in terms of two vector potentials $A^\mu$ and $B^\mu$, and then construct an action as a functional of those two potentials which exhibits the
SL(2,\textbf{R}) invariance in a manifest way. The
SL(2,\textbf{R}) transformation of the potential $A^\mu$ and $B^\mu$ should be provided by the formula (\ref{eq14}).

Since the  equations of motion in (\ref{eq4a2})-(\ref{eq4b2}) have the same form as in the linear case, the curl of $A^\mu$ is  simply related to the filed strengths $G^{\mu\nu}$ as in (\ref{eq18}) and the curl of $B^\mu$ is  simply related to the tensor field $H^{\mu\nu}$ as in (\ref{eq19}). 
This implies that (\ref{eq20a}) is still valid, but (\ref{eq20b}) should be replaced, because of the
complicated relation (\ref{Hfield}) connecting $H^{\mu\nu}$ to $G^{\mu\nu}$, by
\bea
 \frac{e^{-\phi}(n{\cdot}{^{*\!}G})+ \frac{1}{4}e^{-2\phi}(G{^{*\!}G})(n{\cdot} G)}{\sqrt{X}} +a (n{\cdot} G) \ (\!\!&{=}&n{\cdot} {^{*\!}H} \ )\nn \\
 &{=}& n{\cdot} (\partial{\wedge} B) , \label{eqGdotn}
  \eea
where
\bea
X\equiv{1-\frac{1}{2}e^{-\phi}(\, ^{*\!}G)^2-\frac{1}{16}e^{-2\phi}(G ^{*\!}G)^2}. \label{X}
\eea
Instead of trying to solve the nonlinear equation (\ref{eqGdotn}) directly for $n{\cdot} ^{*\!}G$, we 
will below proceed regarding the above quantity $X$ as some given quantity; the appropriate form for the quantity $X$ can be obtained later from the consistency of various equations we have.

Multiplying  (\ref{eqGdotn}) by $n{\cdot} G$  and using the identities
\bea
(G)^2=2 \left( (n{\cdot} G)^2- (n{\cdot} {^{*\!}G})^2\right), \qquad (G{^{*\!}G})=4 (n{\cdot} G){\cdot}(n{\cdot}{^{*\!}G}) \label{idG}
\eea
derived with the help of the  Zwanziger identity (\ref{eq21}),   we find
\bea
\frac{e^{-\phi}(n{\cdot}G){\cdot}( n{\cdot} {^{*\!}G})}{\sqrt{X}}=\frac{[\scA]{\cdot} [\scB]-a\,[\scA]^2}{1+e^{-\phi} [\scA]^2} . \label{nGG}
\eea
Substituting (\ref{nGG}) into (\ref{eqGdotn}), we can determine $n{\cdot}{\,^{*\!}G}$ in terms of the vector potentials $A_\mu$ and $B_\mu$ (and $X$) as
\bea
n\cdot {^{*\!}G}= e^\phi\sqrt{X}\{[\scB]-Y [\scA]\}\label{n*G}
\eea
with
\bea
Y&=&a+ \frac{[\scA]{\cdot} [\scB]-a\,[\scA]^2}{e^{\phi} + [\scA]^2} \\
&=&\frac{a+e^{-\phi}[\scA]{\cdot}[\scB]}{1+e^{-\phi}[\scA]^2}\label{Yeq}
\eea
The results (\ref{eq20a}) and (\ref{n*G}) for $n{\cdot}{G}$ and $n{\cdot}{\,^{*\!}G}$, respectively,
may be used with the Zwanziger identity to express $G^{\mu\nu}$ in terms of the potentials
$A^\mu$ and $B^\mu$. Inserting such expression into (\ref{X}), we  then obtain 
the following consistency condition for the quantity $X$:
\bea
X&=& 1+e^{-\phi}[\scA]^2-e^\phi X\Big[\scB -Y \scA\Big]^2\nonumber \\
&&-X\Big([\scA]{\cdot} [\scB-Y \scA] \Big)^2 .
\eea
Collecting terms involving $X$ to  one side of this equation, we find
\bea
X^{-1}\left( 1+e^{-\phi}[\scA]^2\right)&=&1+e^{\phi}\Big[ \scB-Y \scA\Big]^2\nonumber \\ 
&&+\Big([\scA]\cdot [\scB-Y \scA]\Big)^2 . \label{Xinv}
\eea
After some algebras,  the quantity $X$ is then given in terms of $A^\mu$ and $B^\mu$
as follows:
\bea
X=\frac{1+e^{-\phi} [\scA]^2}{1+e^{\phi}[\scB-a\, \scA]\cdot[\scB -Y \scA]}\label{Xfin}
\eea

By the above procedure,  $n{\cdot}G$ and $n{\cdot}\, ^{*\!}G$  are fully determined in terms of the vector potentials $A^\mu$ and $B^\mu$. Based on these results, we obtain the following representations for $G^{\mu\nu}$ and $H^{\mu\nu}$:
\begin{subequations}
\bea
G&=&n{\wedge}[\scA]-\frac{1}{\sqrt{\cal M}}\Big(e^\phi+[\scA]^2\Big)\, ^{*\!}\{n{\wedge}[\scB]\}\nonumber \\
&&+\frac{1}{\sqrt{\cal M}}\Big(ae^\phi+[\scA]\cdot[\scB]\Big)\, ^{*\!}\{n{\wedge} [\scA]\} ,   \label{G1} \\
H&=&-\ ^{*\!}\{n{\wedge}[\scB]\}-\frac{1}{\sqrt{\cal M}}\Big(e^\phi a+[\scA]{\cdot}[\scB]\Big) n{\wedge} [\scB] \nonumber \\ 
&&+\frac{1}{\sqrt{\cal M}}  \Big(e^{-\phi}+a^2 e^\phi +[\scB]^2\Big) n{\wedge} [\scA], \label{H1}
\eea
\end{subequations}
where
\bea
{\cal M}
&=&1+e^{-\phi}(1+e^{2\phi}a^2)[\scA]^2+e^{\phi}[\scB]^2 -2 e^{\phi}a[\scA]{\cdot}[\scB]\nonumber \\
&&
+[\scA]^2[\scB]^2 -\Big([\scA]{\cdot}[\scB]\Big)^2\ . \label{rootM}
\eea
To obtain the expressions in (\ref{G1}) and (\ref{H1}), we have used the fact that the two
quantities $X$ and ${\cal M}$ are simply related by
\bea
\sqrt{X}\sqrt{{\cal M}}= 1+ e^{-\phi} [\scA]^2 .
\eea

Now we turn to the construction of  an SL(2,\textbf{R})-invariant action from which the equations of motion (\ref{eq4a2})-(\ref{eq4b2}) can be obtained by considering the variations of the vector potentials $A^\mu$ and $B^\mu$.
Let us start from the action form
\bea
S&=&\int\! d^4x \left[1-\sqrt{X(G)}
-\frac{1}{2}\, ^{*\!}G^{\mu\nu}(\partial_\mu B_\nu-\partial_\nu B_\mu)+\frac{1}{4}
a  \, ^{*\!}G^{\mu\nu}\, G_{\mu\nu}\right] \nn \\
&&+\int\! d^4 x\left[ J^\mu_{(e)}A_\mu(G)+J^\mu_{(m)} B_\mu\right],  \label{biaction}
\eea
which may be regarded as a generalization of the form (\ref{eq30}) (for the linear case) to the present nonlinear model.  
Independent variables are assumed by $^{*\!}G^{\mu\nu}$ and $B^\mu$, with $X(G)$ specified as in (\ref{X}) and  $A^\mu(G)$ by (\ref{eq31}), which is a nonlocal relation.
Following the steps below (\ref{eq31}) in the previous section, one may easily verify that  variations of two independent variables  $B^\mu$ and $^{*\!}G^{\mu\nu}$ yield the right set of equations motion (\ref{eq4a2})-(\ref{eq4b2}).


It is also straightforward to find the corresponding local action 
which has the potentials  $A^\mu$ and $B^\mu$ as independent variables, by following the steps similar to those used in the previous section. But algebraic  calculations are significantly more involved. 
Since the $\sqrt{X}$ in (\ref{biaction}) is already given in (\ref{Xfin}),  let us focus on the last two terms  in the first line of (\ref{biaction}). If the representation (\ref{G1}) is used, it can be cast as
\bea
&&
-\frac{1}{2}{^{*\!}G}^{\mu\nu}(\partial{\wedge} B)_{\mu\nu}+\frac{1}{4} a  \, ^{*\!}G^{\mu\nu}\, G_{\mu\nu} \nonumber \\ 
&& =-\frac{1}{2}\Big([n{\wedge}(n{\cdot} \,^{*\!}G)]+ \, ^{*\!}[n{\wedge}(n{\cdot} G)]\Big)^{\mu\nu} 
\Big(n{\wedge}[\scB]- ^{*\!}\{n{\wedge}[n{\cdot} \,^{*\!}(\partial{\wedge} B)] \}\Big)_{\mu\nu}
\nonumber \\
&&\qquad +a\, (n{\cdot} G)\cdot(n{\cdot}^{*\!}G)   \nonumber \\
&&=-\frac{1}{2} [n{\cdot}(\partial{\wedge} A)]\cdot [n{\cdot}{^{*\!}(}\partial{\wedge} B)]+\frac{1}{2} [n{\cdot}(\partial{\wedge} B)]\cdot [n{\cdot}{^{*\!}(}\partial{\wedge} A)]\nonumber \\
&&\qquad - e^{\phi}\sqrt{X}[\scB-a\ \scA]\cdot [\scB-Y \scA].
\eea
Substituting this into (\ref{biaction}) and working some algebras out, we obtain the desired 
Zwanziger-type local action
\bea
S&=&\int\! d^4x \Big[ 1-\frac{1}{2} (n\cdot[\partial{\wedge} A])\cdot (n\cdot{^{*\!}[}\partial{\wedge} B])+\frac{1}{2} (n\cdot[\partial{\wedge} B])\cdot (n\cdot{^{*\!}[}\partial{\wedge} A])\nn \\
&&\qquad \qquad -\sqrt{\cal M}+J_{(e)}^\mu A_\mu+J_{(m)}^\mu B_\mu\Big] \label{BI2}
\eea
with the quantity ${\cal M}$, as a function of the potentials $A^\mu$ and $B^\mu$,  given by 
(\ref{rootM}).
Notice that small field approximation of this Born-Infeld action in (\ref{BI2}) precisely becomes the action in (\ref{eq34}).

As in the case of linear theory, using the complex scalar $\tau=a+i e^{-\phi}$, it is possible to present the  action (\ref{BI2}) in the form
\bea
S&=&\int\! d^4x \Big[ \frac{1}{2} {\rm Im}\left\{e^{\phi}\left(n{\cdot}[\tau\partial{\wedge} A-\partial{\wedge} B]\right)\cdot \left(n{\cdot}\,^{*\!}[\bar{\tau}\partial{\wedge} A-\partial{\wedge} B]\right)\right\}+1-\sqrt{\cal M}\Big]\nn \\ 
&&\quad +\int\! d^4x \Big[J_{(e)}^\mu A_\mu+J_{(m)}^\mu B_\mu\Big]  \label{BI3}
\eea
with the quantity ${\cal M}$ also expressed as
\bea
{\cal M}&=&1+{\rm Re}\left\{e^{\phi}\left(n{\cdot}[\tau\partial{\wedge} A-\partial{\wedge} B]\right)\cdot \left(n{\cdot} [\bar{\tau}\partial{\wedge} A-\partial{\wedge} B]\right)\right\} \nonumber \\
&&+[\scA]^2[\scB]^2 -\Big([\scA]{\cdot}[\scB]\Big)^2. \label{rootM2}
\eea
One may readily see that the action in (\ref{BI3}) reduces to the  action (\ref{eq35}) of the linear theory if a small field approximation is taken. 
Manifest SL(2,\textbf{R}) invariance of the action (\ref{BI3}) can also be demonstrated immediately. Obviously the terms which are quadratic in $A$ and $B$ from
 (\ref{BI3}) and (\ref{rootM2}) are invariant under the SL(2,\textbf{R}) transformation.
Further, the expression in the second line of (\ref{rootM2}) is separately invariant when the potentials transform as in (\ref{eq14}): to see this, note that, under the transformation (\ref{eq14}),
\bea
&&[n{\cdot}(\partial{\wedge} A')]^2[n{\cdot}(\partial{\wedge} B')]^2 -\Big([n{\cdot}(\partial{\wedge} A')]\cdot[n{\cdot}(\partial{\wedge} B')]\Big)^2\nonumber \\
&&=\left\{[\scA]^2[\scB]^2 -\Big([\scA]{\cdot}[\scB]\Big)^2\right\}(pq-rs)^2
\eea
while  the condition $pq-rs=1$ holds for any SL(2,\textbf{R}) transformation.

One can verify (after a straightforward, albeit somewhat lengthy, algebra) that the equations of motion
(\ref{eq4a2})-(\ref{eq4b2}), with $G^{\mu\nu}$ and $H^{\mu\nu}$ represented by the expressions in 
(\ref{G1}) and (\ref{H1}),  respectively, are precisely the stationary conditions with our action
(\ref{BI2}) or (\ref{BI3}) under the variation of the potentials $A^\mu$ and $B^\mu$. Another point
is: based on (\ref{eq4a2})-(\ref{eq4b2}) and representations (\ref{G1}) and (\ref{H1}), it is possible
to deduce the relations which are identical in forms with (\ref{eq37}) and (\ref{eq38}). Hence
the relations of the previous section (see (\ref{eq18}) and (\ref{eq19}))
\begin{subequations}
\bea
G^{\mu\nu}=(\partial{\wedge} A)^{\mu\nu}+(n{\cdot}\partial)^{-1}\, ^{*\!}(n{\wedge} J_{(m)})^{\mu\nu}, \label{G2} \\
H^{\mu\nu}=
-^{*\!}(\partial{\wedge} B)^{\mu\nu}-(n{\cdot}\partial)^{-1}(n{\wedge} J_{(e)})^{\mu\nu} \label{H2}
\eea\end{subequations}
are also valid in this nonlinear theory. This means that, even in the context of Born-Infeld theory, the Lorentz 4-force structure that one may expect on the basis of minimal couplings would be
of the form given in (\ref{eq40}). Moreover, in free space, one can see from (\ref{G2}) and (\ref{H2})
and the relation (\ref{Hfield}) that $\partial{\wedge} B$ and $\partial{\wedge} A$
are not independently given, i.e., the theory contains only one electromagnetic potential associated with
physical `photon'.

\section{Summary and Discussions}

In this work we have presented the Zwanziger-type formalism, based on two four-potentials, for linear and nonlinear electromagnetic field theories generalized in such a way that they may possess maximal duality
symmetry, i.e., the SL(2,\textbf{R}) invariance.
The duality transformation is realized in a very natural manner by the variables of this formalism, and our action---a local functional of potentials---is manifestly invariant under this transformation. 
In the context of quantum field theory with electric and magnetic sources, only the SL(2,\textbf{Z})
subgroup from the classical SL(2,\textbf{R}) duality remains as a surviving duality symmetry.
(Note that the two scalar fields in our theory, representing axion and dilaton, respectively, may
be taken by constants or spacetime-dependent backgrounds or dynamical variables, depending on the problem one is interested in).
The maximally duality-symmetric action constructed by us could serve as a useful basis for more detailed investigations of various quantum-field-theoretic aspects; for instance, the general quantization procedure \cite{cit15,cit16}, problems in perturbative renormalization \cite{cit16}, and chiral anomaly due to dyonic Dirac fields \cite{cit17}, etc. Also interesting would be the related study
on the phase structure of gauge theories in the light of electromagnetic  duality \cite{cit23}. One might also look for a supersymmetric extension of our duality-symmetric action.

Here, we would like to add a comment as regards the equation form in (\ref{eq4a})-(\ref{eq4b}), which was chosen to define our SL(2,\textbf{R}) duality-invariant system. Instead of this form, one might
consider the set
\begin{subequations}
\bea
\partial_\nu(\, ^{*\!}F^{\mu\nu}+\tilde{b} F^{\mu\nu})=J^\mu_{(m)}, \label{mag}\\
\partial_\nu(F^{\mu\nu}-{b} \, ^{*\!}F^{\mu\nu})=J^\mu_{(e)}, \label{elec}
\eea\end{subequations}
which contains  axion-like scalar $b(x)$  together with {\it dual-axion-like} scalar $\tilde{b}(x)$, and try to find
a Zwanziger-like action leading to these equations of motion. No new analysis is required in this case since (\ref{mag}) and (\ref{elec}) can be viewed as a rewriting of our equations (\ref{eq4a}) and (\ref{eq4b}) upon making the connection
\bea 
F^{\mu\nu}-\tilde{b}\, ^{*\!}F^{\mu\nu}&=&G^{\mu\nu}, \quad \, ^{*\!}F^{\mu\nu}+\tilde{b}\, F^{\mu\nu}=\, ^{*\!}G^{\mu\nu}, \nn\\ 
\frac{1+b\tilde{b}}{1+\tilde{b}^2}&=&e^{-\phi}, \qquad\qquad \frac{b-\tilde{b}}{1+\tilde{b}^2}=a.
\eea
(The Lorentz force law that goes with the equations (\ref{mag})-(\ref{elec}) may be inferred
by applying this connection to (\ref{eq5})). With the system (\ref{mag})-(\ref{elec}),
the filed strengths $F^{\mu\nu}$ may thus be expressed using the four-potentials $A^\mu$ and $B^\mu$ as
\bea
F&=&\frac{1}{1+\tilde{b}^2}(\tilde{b}\, ^{*\!}G+G) \nn \\
&=&\frac{1}{1+b \tilde{b}}\Big(n{\wedge}[n{\cdot}(\partial {\wedge}A)+\tilde{b}\,n{\cdot}(\partial {\wedge}B)]
-\, ^{*\!}\{(n{\wedge}[n{\cdot}(\partial {\wedge}B)-{b}\,n{\cdot}(\partial {\wedge}A)]\}\Big),
\eea
where, for the second expression, we  made use of the representations (\ref{eq22}) and (\ref{eq23}).
Also the Zwanziger-like action corresponding to the equations of motion 
(\ref{mag})-(\ref{elec}) is simply our expression (\ref{eq34}) with $e^{-\phi}$ and $a$
now given in terms of $b$ and $\tilde{b}$. This yields
 \bea
 S&=&-\frac{1}{2}\int\! d^4 x \Big\{ [n{\cdot}(\partial{\wedge} A)]{\cdot}[n{\cdot}\, ^{*\!}(\partial{\wedge} B)]- [n{\cdot}(\partial{\wedge} B)]{\cdot}[n{\cdot}\, ^{*\!}(\partial{\wedge} A)] \nn \\
 &&\qquad \qquad\quad+ \frac{1}{1+b\tilde{b}}\Big([n{\cdot}(\partial{\wedge} A)+\tilde{b}\, n{\cdot}(\partial{\wedge} B)]^2+[n{\cdot}(\partial{\wedge} B)-{b}\, n{\cdot}(\partial{\wedge} A)]^2\Big)\Big\} \nn \\
 &&\qquad
+\int\! d^4 x\left[ J^\mu_{(e)}A_\mu+J^\mu_{(m)} B_\mu\right]. \label{eq80}
\eea
This has a structure symmetrical in $(A,B)$ and $(b,\tilde{b})$. But the SL(2,\textbf{R}) duality
symmetry is more transparent when one puts the equations of motion by the form (\ref{eq4a})-(\ref{eq4b}), i.e., all modifications to that of the usual Maxwell theory are arranged to  enter through
the Maxwell equation involving the electric source.
With quantum effects taken into account, this advantage in discussing the duality symmetry will become more important.

Also, in the literature, there exists an alternate form of duality-symmetric
action, originally due to  Schwarz and Sen for the Maxwell action \cite{cit14} and later
extended to the Born-Infeld action by Berman \cite{Berman}.
This other formulation appears superficially similar to that of Zwanziger, 
as both rely on a particular direction $\hat n$ to separate the field strengths
into two parts; on the first look, one can easily be misled to believe 
that a certain redefinition of the variables would bring one to the other.
The role of field strengths along $\hat n$  
and orthogonal to $\hat n$  are opposite in these two
formulations, and we found no simple way to relate these two approaches.
The differences may also have something to to with the mysterious factor 1/2 
that appears
in an extension of Schwarz-Sen action by Deser {\it et al.}\cite{Deser}.
This factor of $1/2$ in the minimal
couplings is needed to make sure that the final equation
of motion come out correctly, i.e., with no such factor  $1/2$. But, as we have seen, Zwanziger's
formulation requires no such extra factor in the action.

({\bf Note added}) After completion of this work, we are told by Prof. Sorokin that the Schwarz-Sen action and the Zwanziger action are dual to each other, as shown in Ref.\cite{sor98}, by considering the dualization with a Lorentz covariant form of the duality-symmetric Maxwell action developed in \cite{sor95}. We are very grateful to him
for this information.


\section*{Acknowledgments}
We are very grateful to Prof. Piljin Yi for illuminating discussions
on the Schwarz-Sen approach and Born-Infeld electrodynamics.
We also acknowledge useful discussions with Prof. Kimyeong Lee. The work of CL was partly performed while he was visiting the Korea Institute for Advanced Study for the year 2012/2013, and he is grateful to Prof. D. Kim for making this visit possible.
HM's work was supported by the National Research Foundation of Korea (NRF) funded by the Ministry of Education (No 2010-0011223).

\newpage

\end{document}